\documentclass[11pt]{article}
\usepackage[left=1in,top=1in,right=1in,bottom=1in,head=.1in,nofoot]{geometry}

\usepackage{setspace,url,bm,amsmath} 
\usepackage{float}
\usepackage{caption}
\usepackage{subcaption,siunitx,booktabs}

\usepackage{multirow}
\usepackage{titlesec} 
\usepackage{rotating}
\titlelabel{\thetitle.\quad} 
\titleformat*{\section}{\bf\large}

\usepackage{graphicx} 
\usepackage{bbm}
\usepackage{latexsym}
\usepackage{color}
\usepackage{authblk}
\usepackage{amsthm}
\usepackage{amsfonts}
\usepackage{bm}
\usepackage{comment}
\usepackage{appendix}

\newtheorem{theorem}{Theorem}
\newtheorem{proposition}{Proposition}
\newtheorem{lemma}{Lemma}

\theoremstyle{definition}

\theoremstyle{remark}
\newtheorem{remark}{Remark}[section]

\usepackage{natbib} 
\bibpunct{(}{)}{;}{a}{}{,} 
\usepackage{enumitem}

\usepackage{etoolbox} 
\apptocmd{\sloppy}{\hbadness 10000\relax}{}{} 

\def\T{\textnormal{T}}
\def \E{\mathbb{E}}
\def\var{\textnormal{Var}}
\def\cov{\textnormal{Cov}}

\def \yio{y_i^{\mathrm{obs}}}

\def \yo{{\mathbf y}^{\mathrm{obs}}}
\def \lambdar{{\bm \lambda}_{\text{R}}}
\def \lambdag{{\bm \lambda}_{\text{G}}}
\def \lambdai{{\bm \lambda}_{\text{I}}}
\def \lambdarg{{\bm \lambda}_{\text{RG}}}
\def \lambdari{{\bm \lambda}_{\text{RI}}}
\def \lambdagi{{\bm \lambda}_{\text{GI}}}
\def \lambdargi{{\bm \lambda}_{\text{RGI}}}
\def \lambdarj{{\bm \lambda}_{\text{R}j}}

\def \taul{\tau_{\ell}}
\def \taulest{\widehat{\tau}_{\ell}}
\def \etar{\eta_{\text{R}}}
\def \thetar{\theta_{\text{R}}}
\def \taur{\tau_{\text{R}}}
\def \taug{\tau_{\text{G}}}
\def \taui{\tau_{\text{I}}}
\def \taurg{\tau_{\text{R}\circ\text{G}}}
\def \tauri{\tau_{\text{R}\circ\text{I}}}
\def \taugi{\tau_{\text{G}\circ\text{I}}}
\def \taurgi{\tau_{\text{R}\circ\text{G} \circ \text{I}}}
\def \taurind{\tau_{i,\text{R}}}
\def \taugind{\tau_{i,\text{G}}}
\def \tauiind{\tau_{i,\text{I}}}
\def \taurgind{\tau_{i,\text{R}\circ\text{G}}}
\def \tauriind{\tau_{i,\text{R}\circ\text{I}}}
\def \taugiind{\tau_{i,\text{G}\circ\text{I}}}
\def \taurgiind{\tau_{i,\text{R}\circ\text{G} \circ \text{I}}}

\newcommand{\taurunit}[1]{\tau_{#1,\text{R}}}
\newcommand{\taugunit}[1]{\tau_{#1,\text{G}}}
\newcommand{\tauiunit}[1]{\tau_{#1,\text{I}}}
\newcommand{\taurgunit}[1]{\tau_{#1,\text{R}\circ\text{G}}}
\newcommand{\tauriunit}[1]{\tau_{#1,\text{R}\circ\text{I}}}
\newcommand{\taugiunit}[1]{\tau_{#1,\text{G}\circ\text{I}}}
\newcommand{\taurgiunit}[1]{\tau_{#1,\text{R}\circ\text{G} \circ \text{I}}}

\def \tauFP{\bm{\tau}^{\text{FP}}}
\def \tauSP{\bm{\tau}^{\text{SP}}}

\title{\bf Analysis and sample-size determination for $2^K$ audit experiments with binary response and application to identification of effect of racial discrimination on access to justice}

\author{Nicole Pashley*, Brian Libgober**, Tirthankar Dasgupta*}

\affil[*]{Department of Statistics, Rutgers University}
\affil[**]{Department of Political Science, Northwestern University}

\begin{document}

\date{}

\doublespacing

\maketitle


\begin{abstract}
Social scientists have increasingly turned to audit experiments to investigate discrimination in the market for jobs, loans, housing and other opportunities.  In a typical audit experiment, researchers assign ``signals'' (the treatment) to subjects at random and compare success rates across treatment conditions. In the recent past there has been increased interest in using randomized multifactor designs for audit experiments, popularly called factorial experiments, in which combinations of multiple signals are assigned to subjects. 
Although social scientists have manipulated multiple factors like race, gender and income, the analyses have been mostly exploratory in nature. 
In this paper we lay out a comprehensive methodology for design and analysis of $2^K$ factorial designs with binary response using model-free, randomization-based Neymanian inference and demonstrate its application by analyzing the audit experiment reported in \textit{Getting a Lawyer While Black} (Libgober, 2020). Specifically, we integrate and extend several sections of the randomization-based, finite-population literature for binary outcomes, including sample size and power calculations, and non-linear factorial estimators, extending results.
\end{abstract}

\section{Introduction}
\label{sec:intro}

Social scientists have increasingly turned to audit experiments (or correspondence experiments) to investigate discrimination in the market for jobs, loans, housing and other opportunities. \cite{Baert2018} identifies over ninety audit experiments conducted between 2005 and 2016.
The literature discussing the interpretation, analysis, and implications of these experiments is even larger \citep{Gaddis2018a}.
In a typical audit experiment, researchers assign ``signals'' (the treatment) to applications at random and compare success rates across treatment conditions.
The signal provides information about the latent traits of the applicant, such as their race, gender, or other demographic information.
Crucially, the applications are kept absolutely identical except for the ``signal'' which reveals the experimental treatment condition. For example, in the seminal study of \cite{Emily2004} on discrimination in labor market, the name atop the resume was ``Emily,'' ``Lakisha,'' or some name similarly suggestive of race.\footnote{It is important to note that in this experiment, researchers manipulate the \textit{presentation} of race, by varying the name attached to the resume. Names such as Lakisha or Emily are highly correlated with ``race'', at least as recorded on birth or other governmental records, and therefore constitute a strong signal about applicant race. Thus, in this case, the trait or attribute itself is not manipulated but rather a signal of it is presented to the participant.}

A key issue that often raises questions about inference drawn from audit experiments as above is confounded signaling of latent traits like race. Such confounding arises because names that are predictive of race are also correlated with myriad other factors, importantly socio-economic status. Indeed, \cite{Fryer2004} show a strong relationship between a child having a racially distinctive name and whether their mother had health insurance at the time of birth. Names may also convey information about an individual's age, nationality, religion, and still other factors. 

A design-based approach to address the challenge arising from confounded signaling is to use \emph{randomized multifactor designs}, popularly called \emph{factorial experiments}. In a factorial experiment, multiple interventions or level combinations of factors are applied simultaneously on a population of experimental units or subjects (e.g., lawyers) using a randomized assignment mechanism. The act of randomization permits researchers to give causal interpretations to observed associations \citep{imbens2015book}. In addition to letting the experimenter identify the ``main'' or marginal effect of each factor, such designs also permit assessment of interactions among factors. Such interactions can be meaningful in the context of discrimination experiments. For example, how different is the effect of race signaling on access to lawyers for people belonging to low income and high income groups? 

Although previous audit studies have manipulated multiple factors \citep[for example,][varied signals for race, gender and income]{Libgober2020}, the analyses have been mostly exploratory in nature, not involving use of the recently developed tools and techniques for drawing design-based causal inference from such designed experiments. Design-based or randomization-based inference is a useful methodology for drawing inference on causal effects of treatments from factorial experiments in a finite-population setting \citep[e.g.,][]{Freedman2006, Freedman2008a}.
A major advantage of such inference is that it applies even if the experimental units are not randomly sampled from a larger population, which is the case in most social science experiments \citep{Abadie2020, Olsen2013} and is true for our experimental setting. Further, the analysis does not require any additional assumptions if the outcome is binary, which is also the case here. 

Experiments with $K$ factors each at two levels are called $2^K$ factorial experiments \citep{BHH2005, WandH2009}. While $2^K$ factorial experiments are used extensively in industrial experiments and increasingly in biomedical experiments, their applications in social and behavioral experiments have remained limited. In this paper, we will \emph{use and extend} recent developments on design-based causal inference for factorial designs with binary responses that revolve around the potential outcomes or counterfactuals framework, introduced by \cite{Neyman1923} and popularized by \cite{rubin_1974}. There has been a recent increase of interest in the design and analysis of factorial experiments using the potential outcomes framework following \cite{DPR2015}. 

Our main contribution is to lay out a comprehensive methodology for analysis of $2^K$ factorial designs with binary response using the model-free, randomization-based asymptotic Neymanian inference and demonstrate its application by re-analyzing the experiment reported in \cite{Libgober2020}. Specifically, we synthesize and extend the results from prior work on experiments with binary outcomes \citep{dingdasgupta2016, Lu2019, lu2019improved} for $2^K$ experiments for any integer $K$ by (i) developing a methodology for determining the sample size based on the level of power desired to identify active causal effects for future studies, (ii) exploring the performance of both asymptotic and finite-corrected methods in the finite-population, randomization-based setting through simulations, and (iii) defining non-linear factorial effects for binary responses, that can be considered as generalizations of the risk ratio and risk odds ratio, and proposing methods for their asymptotic inference.
Throughout we focus on nuances of doing finite-population inference which sets it apart from super-population inference.

The paper is organized as follows: 
We complete this section with a description of the motivating example which we will use to anchor the framework and methodology throughout the paper.
Section \ref{sec:notation} defines the causal estimands in terms of potential outcomes and observed data. 
Section \ref{sec:Neyman} describes Neymanian inference of factorial effects and a procedure for sharpening the inference, illustrating the procedure with the experiment conducted by \cite{Libgober2020}. Section \ref{sec:power_sample} details a procedure for calculating the desired sample size based on the desired power of hypothesis testing. 
Section \ref{sec:other_estimands} defines non-linear factorial estimands for binary responses and proposes methods for their asymptotic inference. 
Some concluding remarks, discussion and opportunities of future work are presented in Section \ref{sec:discussion}.

\cite{Libgober2020} reported a pilot study to determine whether black individuals and white individuals face equal barriers in acquiring legal services from lawyers. The experimental units were lawyers in private practices, selected from the California bar directory. Requests for representation from potential clients who claimed to be wrongfully accused of driving under the influence (DUI) were sent to lawyers by email. In these emails, race was signaled through the use of a name highly correlated with a specific race - black (denoted by 0) or white (denoted by 1). Two other factors - income (low, denoted by 0 and high, denoted by 1) and gender (female, denoted by 0 and male, denoted by 1) - were also signaled in the emails. Thus, each lawyer received an email that incorporated one of the eight possible level combinations of the three treatments $000$, $001$, $010$, $011$, $100$, $101$, $110$ and $111$.  96 lawyers were selected for the study, and each of the eight treatment combinations were assigned to 12 lawyers using a completely randomized assignment mechanism. The primary outcome of interest was binary taking value 1 if the lawyer replied in any fashion at all and 0 otherwise. 

\section{Set up and formulation of the problem in the potential outcomes framework} \label{sec:notation}

\subsection{Potential outcomes and estimands} \label{ss:POT}

Here, we introduce some key definitions and notation from \cite{DPR2015}. Consider a $2^K$ experiment with $N$ units, in which the levels of each of the $K$ factors are denoted by 0 and 1. Let the treatment combinations be arranged lexicographically starting with $(0 \ 0 \ \ldots \ 0)$ and ending with $(1 \ 1 \ \ldots \ 1)$. Thus, for example, in the $2^3$ experiment described above, the eight treatment combinations starting with $(0 \ 0 \ 0)$ and ending with $(1 \ 1  \ 1)$ as shown in Table \ref{tab:tr_comb} are numbered as $j = 1, 2, \ldots, 8$, respectively.

\begin{table}[htbp]
\centering  \small
\caption{Treatment combinations in a $2^3$ experiment} \label{tab:tr_comb}
\begin{tabular}{c|cccccccc}
Treatment number ($j$)  & 1 & 2 & 3 & 4 & 5 & 6 & 7 & 8 \\ \hline
Level combination of (Race, gender, income) & 000 & 001 & 010 & 011 & 100 & 101 & 110 & 111 \\ \hline
\end{tabular}
\end{table}

For $i = 1, \ldots, N$, under the Stable Unit Treatment Value Assumption or SUTVA \citep{rubin1980}, the $i$th unit has $J = 2^K$ potential outcomes, $Y_i(1), \ldots, Y_i(J)$, corresponding to the $J$ treatment combinations. Let $\mathbf{Y}_i$ denote the $J \times 1$ vector of potential outcomes for unit $i$.  For unit $i$, the unit-level main effect of factor $k\in\{1, \ldots, K\}$ is defined as the difference between the averages of potential outcomes for unit $i$ for which the levels of factor $k$ are at levels 1 versus 0.
Let $\bm{z}_j \in \{0,1\}^K$ be the binary representation of treatment $j$ corresponding to the level of each factor, such that $\bm{z}_{j,k} \in \{0,1\}$ represents the $k$th element of $\bm{z}_j$ which is the treatment level of factor $k$.
For example, in Table~\ref{tab:tr_comb} we see that $\bm{z}_2 = 001$ and $\bm{z}_{2,1} = 0$.
Then the main effect of factor $k$ for unit $i$ is
\[\tau_{i,k} = \frac{1}{2^{K-1}}\sum_{j: \bm{z}_{j,k} =1}Y_i(j) - \frac{1}{2^{K-1}}\sum_{j: \bm{z}_{j,k} =0}Y_i(j).\]
More generally, we can write this using a contrast vector (i.e., a vector whose elements sum to 0 but are not all zero), $\bm{\lambda}_k$, where the $j$th entry of this vector is  $\bm{\lambda}_{k,j} = 2\bm{z}_{j,k}-1$.
In other words, $\bm{\lambda}_{k,j}$ is 1 if factor $k$ is at level 1 in the $j$th treatment and is -1 if factor $k$ is at level 0 in the $j$th treatment.
Then $\tau_{i,k} = 2^{-(K-1)}\bm{\lambda}_k\bm{Y}_i$.

In our application, $N = 96$ and the $i$th unit (lawyer) has 8 potential outcomes $Y_i(1), \ldots, Y_i(8)$, corresponding to the 8 treatment combinations of Table \ref{tab:tr_comb}. The unit-level main effect of the race signal, shown in the sixth column (titled R) of Table \ref{tab:PO} is thus 
\begin{equation}
\taurind = \sum_{j=5}^8 Y_i(j)/4 - \sum_{j=1}^4 Y_i(j)/4 = (1/4) \lambdar^{\T} \mathbf{Y}_i, \label{eq:taurind} 
\end{equation}
where $\lambdar = (-1, \ -1, \ -1,  \ -1, \ +1, \ +1, \ +1, \ +1)^{\T}$. Similarly we can define unit-level main effects for the gender and income as $\taugind = (1/4) \lambdag^{\T} \mathbf{Y}_i$ and $\tauiind = (1/4) \lambdai^{\T} \mathbf{Y}_i$, respectively, where $\lambdag =(-1, \ -1, \ +1,  \ +1, \ -1, \ -1, \ +1, \ +1)^{\T}$ and  $\lambdai =(-1, \ +1, \ -1,  \ +1, \ -1, \ +1, \ -1, \ +1)^{\T}$ are the corresponding contrast vectors. 

\begin{table} [htbp]
\centering \footnotesize
\caption{Potential outcomes and estimands} \label{tab:PO}
\renewcommand\arraystretch{1.15} 
\begin{tabular} {c|c|c|c|c|c|c|c|c|c|c|c}
 Unit      & \multicolumn{4}{c|}{Level Combinations} & \multicolumn{7}{c}{Factorial effects}  \\
($i$) & 1: $000$ & 2: $001$ & $\cdots$ & 8: $111$ & R & G & I & RG & RI & GI & RGI \\ \hline
1              &  $Y_1(1)$     &   $Y_1(2)$  &  $\cdots$ & $Y_1(8)$  & $\taurunit{1}$ & $\taugunit{1}$ & $\tauiunit{1}$ & $\taurgunit{1}$ & $\tauriunit{1}$ & $\taugiunit{1}$ & $\taurgiunit{1}$ \\
$\vdots$   &  $\vdots$     &   $\vdots$  &  $\vdots$ &  $\vdots$ & $\vdots$ & $\vdots$  & $\vdots$  & $\vdots$ & $\vdots$ & $\vdots$ & $\vdots$ \\
96             & $Y_{96}(1)$     &   $Y_{96}(2)$  &  $\cdots$ & $Y_{96}(8)$ & $\taurunit{96}$ & $\taugunit{96}$ & $\tauiunit{96}$  & $\taurgunit{96}$ & $\tauriunit{96}$ & $\taugiunit{96}$ & $\taurgiunit{96}$\\
\hline \hline
Average & $\overline{Y}(1)$ & $\overline{Y}(2)$ &  $\cdots$ & $\overline{Y}(8)$ & $\taur$ &  $\taug$ & $\taui$ & $\taurg$ & $\tauri$ & $\taugi$ & $\taurgi$ \\ 
            & $=P_1$ & $=P_2$ &  $\cdots$ &  $=P_{8}$ & & & & & & & \\ \hline \hline 
Variance & $S_1^2$ & $S_2^2$  & $\cdots$ & $S^2_8$ & $S^2_{\taur}$ & $S^2_{\taug}$ & $S^2_{\taui}$ & $S^2_{\taurg}$ & $S^2_{\tauri}$ & $S^2_{\taugi}$ & $S^2_{\taurgi}$ \\ \hline
\end{tabular}
\end{table}

Proceeding along the lines of \cite{DPR2015}, interactions can be defined as contrasts of the form $2^{K-1} {\bm \lambda}^{\T} \mathbf{Y}_i$, where the contrast vector ${\bm \lambda}$ for any interaction can be derived by element-wise multiplication of the contrast vectors of the corresponding main effects, for factors involved in the interaction.
In addition to interactions, it's useful to define the average of all potential outcomes as 
\[\tau_{i,0} = \frac{1}{2^{K}}\sum_{j=1}^JY_i(j).\]
We can then collect all factorial effects into a vector $\bm{\tau}_i = (2\tau_{i,0}, \tau_{i,1},\dots,\tau_{i,K}, \tau_{i,1\circ2}, \dots, \tau_{i,1\circ 2 \dots \circ K})$, where the $\circ$ notation separates factors involved in the interaction. We will denote the $J-1$ unit-level factorial effects by $\tau_{i \ell}$, $i=1, \ldots, N$ and $\ell = 1, \ldots, J-1$, where $\ell$ is an index over all $J-1$ factorial effects.

In our example, for unit $i$, we can define three two-factor interactions $\taurgind$ (race $\times$ gender), $\tauriind$ (race $\times$ income) and $\taugiind$ (gender $\times$ income) and one three-factor interaction $\taurgiind$ (race $\times$ gender $\times$ income). 
The contrast vector for the three-factor interaction, $\lambdargi$, is obtained by element-wise multiplication of $\lambdar$, $\lambdag$ and $\lambdai$ and is $(-1 \ +1 \ +1  \ -1 \ +1 \ -1 \ -1 \ +1)^{\T}$. The seven unit-level factorial effects are shown in the last seven columns of Table \ref{tab:PO}. Let 
\begin{equation}
{\bm \tau}_i = \left( 2\tau_{i,0}, \taurind, \taugind, \tauiind, \taurgind, \tauriind, \taugiind, \taurgiind \right)^{\T}, \label{eq:unitlevelvec}
\end{equation}
denote the $8 \times 1$ vector of the seven unit-level factorial effects, with the first element being twice the average of all 8 potential outcomes for unit $i$.

The finite-population level counterpart of each unit-level factorial effect can be defined by averaging the unit-level factorial effects over the $N$ units. 
That is, we drop the $i$ notation and write 
\[ \tau_k = N^{-1} \sum_{i=1}^N \tau_{i,k} =  \bm{\lambda}_k^{\T} \overline{\mathbf{Y}}, \]
as the average main effect for factor $k$ among all $N$ units in the experiment, where $\overline{\mathbf{Y}} = \left( \overline{Y}(1), \ldots, \overline{Y}(J) \right)$ is the vector of the average potential outcomes for each treatment combination.
For example, the average main effect of race is 
$$ \taur = N^{-1} \sum_{i=1}^N \taurind = (1/4) \lambdar^{\T} \overline{\mathbf{Y}}, $$
where the meaning of $\overline{\mathbf{Y}} $ is illustrated in the row of averages of Table \ref{tab:PO}.
Finite-population interactions are defined analogously.

The finite-population level causal estimands are all of these factorial effects -- the main effects and interactions.
In our example, the seven factorial effects are $\taur$, $\taug$, $\taui$, $\taurg$, $\tauri$, $\taugi$, $\taurgi$ where suffixes R, G and I represent race, gender and income, respectively, $\taur$, $\taug$, $\taui$ are the main effects, $\taurg$, $\tauri$ and $\taugi$ represent the two-factor interactions, and $\taurgi$ represents the three factor interaction.
Each factorial effect, shown in the row of averages in Table \ref{tab:PO}, is the average of the corresponding unit-level factorial effects.

Note that because here we focus on binary potential outcomes, the average $\overline{Y}(j)$ actually represents a proportion $P_j$.
In our example, $P_j$ is the proportion of lawyers who would respond to an email based on treatment combination $j$.
Thus, $\overline{\mathbf{Y}} \equiv \mathbf{P} = (P_1, \ldots, P_J)^{\T}$ and each factorial effect can also be expressed as a contrast of $\mathbf{P}$,
\begin{equation}
\taul = \frac{1}{2^{K-1}} \bm{\lambda}_\ell^{\T} \mathbf{P} = \frac{1}{N} \sum_{i=1}^N \tau_{i, \ell}, \label{eq:tau_R} 
\end{equation}
where the index $\ell$ represents the $J-1$ factorial effects (main effects and interactions).
The vector of proportions is shown in the row of averages of Table \ref{tab:PO}.

 Also, note that the last row of Table \ref{tab:PO} shows the variances $S^2_j$'s of the potential outcomes corresponding to treatment combination $j$ with divisor $N-1$, i.e.,
\begin{equation}
S^2_j = \frac{1}{N-1} \sum_{i=1}^N \left(Y_i(j) - \overline{Y}(j) \right)^2 = \frac{N}{N-1} P_j (1-P_j). \label{eq:Ssqj}
\end{equation}

The variance of the unit-level main effects of factor $k$ is 
\begin{equation}
S^2_{\tau_k} = \frac{1}{N-1} \sum_{i=1}^N \left( \tau_{i,k} - \tau_k \right)^2. \label{eq:S^2tau}
\end{equation}
The variance of the unit-level main effects of race, $S^2_{\taur}$, is shown in the last row of Table \ref{tab:PO}.
Variances of unit-level interaction effects are defined analogously.

To obtain an easy representation of the factorial effects in terms of potential outcomes, we define a $J \times J$ matrix, $\bm{L}$, by combining ${\bm \lambda}_0$, the $J \times 1$  vector with all elements equal to one, with the all the other contrast vectors for the factorial effects.

In our example, this results in an $8 \times 8$ matrix as follows: 
\begin{eqnarray}
\bm{L} &=& ({\bm \lambda}_0, \lambdar, \lambdag, \lambdai, \lambdarg, \lambdari, \lambdagi, \lambdargi) \nonumber \\
&=& \left( \begin{array}{rrrrrrrr}
+1  &  -1  &  -1  &   -1  &+1  &+1  &+1  & -1 \\
+1  &  -1  &  -1  &  +1  &+1  & -1  & -1  &+1 \\
+1  &  -1  & +1  &   -1  & -1  &+1  & -1  &+1 \\
+1  &  -1  & +1  &  +1  & -1  & -1  &+1  & -1 \\
+1  & +1  &  -1  &   -1  & -1  & -1  &+1  &+1 \\
+1  & +1  &  -1  &  +1  & -1  &+1  & -1  & -1 \\
+1  & +1  & +1  &   -1  &+1  & -1  & -1  & -1 \\
+1  & +1  & +1  &  +1  &+1  &+1  &+1  &+1 \\
\end{array} \right). \label{eq:matrixL}
\end{eqnarray}
We note that $\mathbf{L}$ is (up to a constant) an orthogonal matrix with $\mathbf{L} \mathbf{L}^\T = \mathbf{L}^\T \mathbf{L} =2^{K} \mathbf{I}_8$, where $\mathbf{I}_J$ denotes the identity matrix of order $J$. 
The linear transform between the vector of average potential outcomes $\overline{\mathbf{Y}}$ and the vector of finite-population level factorial effects $\tauFP$ can be expressed as 
\begin{equation}
\tauFP = \frac{1}{2^{K-1}} \mathbf{L}^{\T} \overline{\mathbf{Y}} =  \frac{1}{2^{K-1}}  \mathbf{L}^{\T} \mathbf{P}. \label{eq:tauFP}
\end{equation} 

\vspace{0.1 in}

\noindent \textbf{A super-population perspective}

\vspace{0.1 in}

While the finite-population perspective does not depend on any hypothetical data generating process for the outcomes, alternative approaches assume that the potential outcomes are drawn from a, possibly hypothetical, super population. Assuming that $\mathbf{Y}_1, \ldots, \mathbf{Y}_N$ are independent and identically distributed random vectors with $E[\mathbf{Y}_i] = {\bm \pi}$, factorial effects at a super-population level are defined as
\begin{equation}
\tauSP = \frac{1}{2^{K-1}}  \mathbf{L}^{\T} {\bm{\pi}}. \nonumber 
\end{equation}
\cite{DL2017SP} discussed the conceptual and mathematical connections between finite- and super-population inference, showing that while the same estimator commonly used to estimate $\tauFP$ unbiasedly is also an unbiased estimator of $\tauSP$, its sampling variances under the two perspectives are different. In the current experiment, if we assume that the 96 lawyers were randomly selected from a larger ``target'' population (e.g., all lawyers in California), and if the goal is to draw inference on such a population, the superpopulation estimand $\tauSP$ will be of interest.

\subsection{Observed outcomes and summary statistics} \label{ss:obs_out}


In a randomized experiment with $N_j$ units assigned to treatment combination $j \in \{1, \ldots, J\}$, only one of the $J$ potential outcomes is observed for unit $i$. This observed outcome is $\yio = Y_i(T_i)$ for $i=1, \ldots, N$, where $T_i \in \{1,\dots,J\}$ is the random treatment assignment for unit $i$ taking value $j$ if unit $i$ receives treatment $j$.
There are $N!/( N_1! \ldots N_J!)$ possible assignments of $N$ units into the $J$ treatment groups such that treatment group $j$ has $N_j$ untis. A completely randomized design picks one of these possible assignments with equal probability. For $j=1, \ldots, J$, let 
$$ n_{j1} = \sum_{i:T_i = j} \yio \  \mbox{and}  \ n_{j0} = \sum_{i:T_i = j} (1-\yio), $$
where $n_{j1} + n_{j0} = N_j$.
We also denote the observed proportion of responses of 1 to treatment $j$ as $p_j = n_{j1}/N_j$. Let $\mathbf{p} = (p_1, \ldots, p_J)$ denote the vector of observed proportions for the $J$ treatment groups. Finally, we define the sample variance of group $j$ as
\begin{equation}
s^2_j = (N_j-1)^{-1} \sum_{i:T_i = j} \left( \yio - p_j \right)^2 = \frac{N_j}{N_j-1} p_j (1-p_j). \label{eq:sj_sq}
\end{equation}

\cite{Libgober2020} used a completely randomized balanced treatment assignment mechanism by randomly dividing the $N=96$ lawyers into $J = 8$ groups of $N_j = 12$ each, where all lawyers in group $j$ received treatment combination $j = 1, \ldots, 8$. The experiment thus generated a $96 \times 1$ vector of observed binary outcomes $\yo$. Here, $p_j = n_{j1}/12$ denotes the observed proportion of lawyers exposed to treatment group $j$ who responded to emails. The observed outcomes from the experiment are summarized in Table \ref{tab:lawyer_outcomes1}, with the last two rows showing the proportions, $p_j$'s, and the sample variances, $s_j^2$'s defined in (\ref{eq:sj_sq}) for $j=1, \ldots, 8$.

\begin{table} [htbp]
\centering \footnotesize
\caption{Summary of observed outcomes from \cite{Libgober2020} experiment} \label{tab:lawyer_outcomes1}
\begin{tabular} {c|c|c|c|c|c|c|c|c}
       & \multicolumn{8}{c}{Treatment combination ($j$)}  \\
Measure & 1: $000$ & 2: $001$ & 3: $010$ & 4: $011$ & 5: $100$ & 6: $101$ & 7: $110$ & 8: $111$ \\ \hline
$N_j$              &  12     &   12  &  12 & 12 & 12 & 12 & 12  & 12   \\ \hline
$n_{j1}$          &  2  & 2 & 2 & 3 & 5 & 2 & 5 & 6 \\ \hline
$n_{j0}$          &  10 & 10 & 10 & 9 & 7 & 10 & 7 & 6 \\ \hline
$p_j$              & 0.167 & 0.167 & 0.167 & 0.250 & 0.417 & 0.167 & 0.417 & 0.500 \\ \hline
$s^2_j$           & 0.1515 & 0.1515 & 0.1515 & 0.2045 & 0.2652 & 0.1515 & 0.2652 & 0.2727 \\ \hline
\end{tabular}
\end{table}

\begin{figure}[ht]
\centering 
\caption{Exploratory analysis: Main effect and two-factor interaction plots for the pilot experiment} \label{fig:MEINTplot}
\includegraphics[width=10cm]{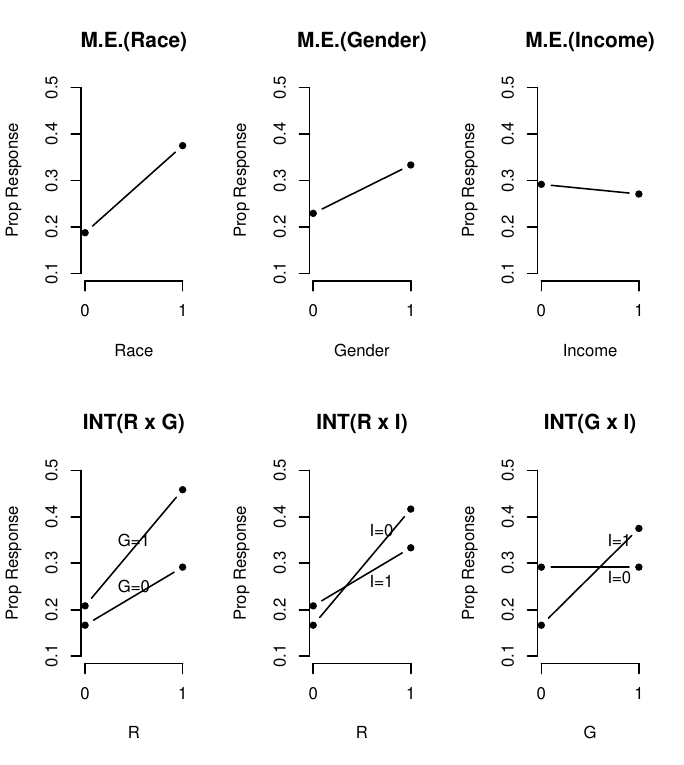} 
\end{figure}

A quick visualization of the observed factorial effects can be provided by the main-effects and two-factor interaction plots \citep{WandH2009}. The main-effect plot of a factor is obtained by plotting the average responses ($p_j$'s in this case) for the two levels of the factor under consideration. The two-factor interaction plot for two factors is obtained by plotting the average responses for the four combinations $00$, $01$, $10$ and $11$ of the two factors under consideration, averaging over the other factors uniformly. Figure \ref{fig:MEINTplot} shows plots of the main effects and two-factor interactions in the example, which suggests that race has the largest main effect, followed by gender. Among the interactions, the $G \times I$ interaction appears to be the strongest. Whereas among emails that appear to come from high-income individuals, there seems to be an increase in responses to males compared to females, such a difference is not noticeable among the emails appearing to come from lower-income individuals.

This exploratory analysis will be followed up by formal asymptotic inference of each factorial effect in the next section to determine which effects are statistically significant at specific levels.

\section{Unbiased estimation and Neymanian asymptotic inference} \label{sec:Neyman}

Note that the sample proportions, $p_j$'s, and sample variances, $s^2_j$'s, defined in the previous section are unbiased estimators of their finite-population counterparts $P_j$'s and $S^2_j$'s, respectively.
Then by substituting the vector of observed proportions $\mathbf{p}$ in place of the population proportions $\mathbf{P}$ in (\ref{eq:tauFP}), we can unbiasedly estimate the vector of factorial effects $\tauFP$ as
\begin{equation}
\widehat{\bm{\tau}} = \frac{1}{2^{K-1}}  \mathbf{L}^{\T} \mathbf{p}. \label{eq:est_tau}
\end{equation}

Using a result from \cite{Lu2016}, the unbiased estimator $\widehat{\tau}$ for a $2^K$ factorial experiment with a completely randomized allocation has covariance matrix
\begin{equation}
\mathbf{V}_{\bm \tau} = \var \left(\widehat{\bm \tau} \right) = \frac{1}{2^{2(K-1)}}  \sum_{j=1}^{J} \frac{S^2_j}{N_j} \widetilde{\bm  \lambda}_j \widetilde{\bm \lambda_j} ^{\T} - \frac{1}{N(N-1)} \sum_{i=1}^N  \left( \bm{\tau}_i  - \tauFP \right) \left( \bm{\tau}_i  - \tauFP \right)^{\T}, \label{eq:cov_matrix}
\end{equation}
where $\widetilde{\bm \lambda}_j$ represents the transpose of row $j$ of the model matrix $\mathbf{L}$ defined in (\ref{eq:matrixL}), ${\bm \tau}_i$ denotes the vector of unit-level factorial effects given by (\ref{eq:unitlevelvec}), $\tauFP$ the vector of finite-population level factorial effects given by (\ref{eq:tauFP}), and $S^2_j$ as defined in (\ref{eq:Ssqj}). Using asymptotic normality of $\widehat{\tau}$ derived using the finite-population central limit theorem \citep[][see Appendix \ref{ss:Appendix1} for formal statement of result in this context]{LiDing2017}, approximate $100(1-\alpha)\%$ confidence intervals for each individual factorial effect $\taul$ can be obtained as 
\begin{equation}
\taulest \pm z_{\alpha/2} \ \sqrt{\var(\taulest)}, \label{eq:CItrue}
\end{equation}
where $z_{\alpha}$ denotes the upper-$\alpha$ point of a standard normal distribution, and $\var(\taulest)$ denotes the appropriate diagonal element of $\mathbf{V}_{\bm \tau}$ defined in (\ref{eq:cov_matrix}).

Because the second term in (\ref{eq:cov_matrix}) involves individual factorial effects and is not identifiable, a conservative ``Neymanian'' estimator of $\mathbf{V}_{\bm \tau}$ is given by
\begin{equation}
\widehat{\mathbf{V}}_{\bm \tau} = \frac{1}{2^{2(K-1)}}  \sum_{j=1}^{J} \frac{s^2_j}{N_j} \widetilde{\bm  \lambda}_j \widetilde{\bm \lambda_j} ^{\T}, \nonumber 
\end{equation}
where $s^2_j$'s are the sample variances defined in (\ref{eq:sj_sq}) and shown in the last row of Table \ref{tab:lawyer_outcomes1}. Each diagonal element of $\widehat{\mathbf{V}}_{\bm \tau}$ equals $2^{-2(K-1)}  \sum_{j=1}^{J} s^2_j/N_j$, and gives a conservative (i.e., positively biased) estimator of the variance of each individual factorial effect.  Replacing  $\sqrt{\var(\taulest)}$ in (\ref{eq:CItrue}) by its estimator
\begin{equation}
\widehat{\mathrm{S.E.}}(\taulest) = \sqrt{\frac{1}{2^{2(K-1)}}  \sum_{j=1}^{J} \frac{s^2_j}{N_j} }, \label{eq:SE}
\end{equation}
one can obtain estimated $100(1-\alpha)\%$ confidence intervals for each individual factorial effect $\taul$ as 
\begin{equation}
\taulest \pm z_{\alpha/2} \ \widehat{\mathrm{S.E.}}
(\taulest). \nonumber 
\end{equation}

To test whether an individual factorial effect, e.g., the finite-population main effect of race, is significantly different from zero, we use the test statistic $T_{\ell} = \taulest/\widehat{\mathrm{S.E.}}(\taulest)$. Let $T_{\ell}^{\mathrm{obs}}$ denote the observed value of the test statistic. The estimated $p$-values for $H_0: \tau_\ell =0$ against one-sided ($\tau_\ell >0$) and two-sided ($\tau_\ell \ne 0$) alternatives are respectively
\begin{eqnarray*}
p_{\mathrm{one-sided}} &=& \mathrm{pr} \left( T_{\ell} \ge T_{\ell}^{\mathrm{obs}} \right) \approx 1 - \Phi(T_{\ell}^{\mathrm{obs}}), \\ 
p_{\mathrm{two-sided}} &=& \mathrm{pr} \left( |T_{\ell}| \ge |T_{\ell}^{\mathrm{obs}}| \right) \approx 2 \left( 1 - \Phi( |T_{\ell}^{\mathrm{obs}}|) \right),
\end{eqnarray*}
where ``$\approx$'' denotes approximately and $\Phi(\cdot)$ denotes the cumulative distribution function (CDF) of a standard normal distribution. 

\begin{remark} \label{rem:conservative}
The inferential procedure for each factorial effect $\tau_\ell$ stated above is ``conservative'' in the sense that the variance (or standard error) of each estimated factorial effect overestimates the true variance (or true standard error) unless the variance of unit-level factorial effect $S^2_{\tau_\ell}$ defined in (\ref{eq:S^2tau}) is zero.
That is, the estimated p-value is expected to be larger than the true p-value.
See Appendix \ref{ss:Appendix2} for formal statement.
How conservative the approximation is will depend on the magnitude of the treatment effect heterogeneity ($S^2_{\tau_\ell}$), with larger heterogeneity implying a more conservative p-value.
However, as noted in \cite{branson2022power}, we usually do not expect the treatment effect heterogeneity to increase in isolation.
Rather increasing effect heterogeneity  typically implies that the variability of potential outcomes under at least one treatment arm is increasing, which may offset some of the conservative impact.
\end{remark}

\begin{remark} \label{rem:one-sided}
The one-sided alternative may be a more appropriate choice for the main effect of race than a two-sided one, because if an average discriminatory effect exists, it is expected to be uni-directional (e.g., emails signaling sender as black being less likely to receive a response than emails signaling sender as white). 
\end{remark}

\begin{table}[ht]
\centering \footnotesize 
\caption{Inference for individual factorial effects} \label{tab:analysis1}
\begin{tabular}{l|rrrcc}
\hline
Factorial effect                                          &  Estimate & S.E.  & Test statistic  &  95\% interval & $p_{\mathrm{two-sided}}$ \\ \hline
Race   & 0.1875  & 0.0917 &  2.0447 & [0.0078,  0.3672] & 0.0409 \\
Gender & 0.1042  & 0.0917 &  1.1363 & [-0.0755, 0.2839] & 0.2558 \\
Income & -0.0208 & 0.0917 & -0.2268 & [-0.2005, 0.1589] & 0.8206 \\
Race $\times$ Gender & 0.0625  & 0.0917 &  0.6816 & [-0.1172. 0.2422] & 0.4955 \\
Race $\times$ Income & -0.0625 & 0.0917 & -0.6816 & [-0.2422, 0.1172] & 0.4955 \\
Gender $\times$ Income & 0.1042 &0.0917 & 1.1363  & [-0.0755, 0.2839] & 0.2558 \\
Race $\times$ Gender $\times$ Income & 0.0625 & 0.0917 & 0.6816 & [-0.1172, 0.2422] & 0.4955  \\ \hline
\end{tabular}
\end{table}

Table \ref{tab:analysis1} summarizes the finite-population inference based on the procedure described in this section. The estimates in the second column are computed by substitution of $\mathbf{p} = (p_1, \ldots, p_8)^{\T}$ from Table \ref{tab:tr_comb} into (\ref{eq:est_tau}), and the S.E. in column 3 is computed by substituting the $s^2_j$'s from Table \ref{tab:tr_comb} and $N_j = r = 12$ for $j=1, \ldots, 8$ into (\ref{eq:SE}). The results reveal that the main effect of race is significant at 5\% level of significance against a two-sided alternative. The next two largest factorial effects are the main effect of gender and the interaction between gender and income.

\subsection{Ensemble-adjusted procedures} \label{ss:IER-EER}

The inferential procedure described in the previous section can be used to construct confidence intervals and assess hypotheses associated with each individual factorial effect. 
For example, a level $\alpha$ rejection rule for the hypothesis $H_0: \taul = 0$ against the two-sided alternative $H_1: \taul \ne 0$ is to reject the null hypothesis if the observed $p$-value $p_{\mathrm{two-sided}}^{\mathrm{obs}} = 2 \left( 1 - \Phi(T_{\ell}^{\mathrm{obs}}) \right)$ is smaller than or equal to $\alpha$, or equivalently, if $|T_{\ell}^{\mathrm{obs}}| \ge z_{\alpha/2}$, where $z_{\alpha}$ denotes the upper $\alpha$-point of a standard normal distribution. This process guarantees that (asymptotically) the type-I error (probability of identifying a factorial effect as active when actually it is not) does not exceed $\alpha$. In the factorial experiment literature, this procedure is typically referred to as the \emph{individual error rate} (IER) control procedure \citep{WandH2009}, also known as the comparison-wise error rate.

When we consider the problem of identifying the active effects among the $J-1$ factorial effects, we encounter the issue of multiple testing due to all $J-1$ null hypotheses of zero average factorial effects being tested simultaneously. In such situations, the probability of incorrectly concluding that at least one effect is active when actually none are active is referred to as the \emph{experiment-wise error rate} (EER). A straightforward approach to controlling EER is to use the so-called Bonferroni correction \citep{WandH2009} or ensemble adjusted $p$-values \citep{RRensemble}. In this approach, each null hypothesis $H_0: \taul=0$, where $\ell = 1, \ldots, J-1$ represents an index for all $J-1$ factorial effects (which include all main effects and interactions), is rejected against the two-sided alternative if the ensemble-adjusted $p$-value 
$$ \min\left\{ 2(J-1)  \left( 1 - \Phi( |T_{\ell}^{\mathrm{obs}}| ) \right), 1 \right\}$$ does not exceed $\alpha$.
An equivalent rejection rule is $|T_{\ell}^{\mathrm{obs}}| \ge z_{\alpha/(2 [J-1])}$, where $T_{\ell} = \taulest/\widehat{\mathrm{S.E.}}(\taulest)$ is the test statistic for the $\ell$-th factorial effect and $T_{\ell}^{\mathrm{obs}}$ is its observed value. 
The Bonferroni procedure is known to be conservative, as shown in Table~\ref{tab:EER}.

\begin{table}[htbp]
\centering \small
\caption{Bonferroni adjusted $p$-values} \label{tab:EER}
\begin{tabular}{|c|c|c|c|c|c|c|c|} \hline
Factorial effect    &  R  & G & I & $R \times G$ & $R \times I$ & $G \times I$ & $R \times G \times I$ \\ \hline
Bonferroni-adjusted & 0.29 & 1.00 & 1.00 & 1.00 & 1.00 & 1.00 & 1.00 \\ \hline
\end{tabular}
\end{table}

As observed by \cite{WandH2009}, in screening experiments, IER is more powerful and preferable because typically many of the factorial effects are negligible, and the EER-adjusted $p$-values are inflated by considering many factorial effects. In the current experiment, only the race factor appears to be active, and use of the EER-adjusted procedure appears to unnecessarily inflate the $p$-value.

\section{Power analysis and sample size calculations for future experiments} \label{sec:power_sample}

We now demonstrate how to conduct a power analysis to assess the sample size requirement for future experiments to be conducted in similar populations.
For similar design-based calculations for two-armed complete randomization and rerandomization, see \cite{branson2022power}.
The power of the two-sided level $\alpha$ test associated with the hypothesis $\taul=0$ for a single factorial effect $\taul$ following the procedure described in Section \ref{sec:Neyman} when the true effect is $\taul^*$ is given by
\begin{eqnarray}
 \beta \left(\taul^* \right) &=& \Pr \left( |T_{\ell}| \ge z_{\alpha/2} | \ \taul = \taul^* \right) = 1 - \Pr \left( \left| \frac{\taulest}{\widehat{\mathrm{S.E.}} (\taulest)} \right| \le z_{\alpha/2} \ \Big| \ \taul = \taul^* \right)  \nonumber \\
   &\approx& 2 - \Phi\left( \frac{\mathrm{S.E.}^{\text{lim}} (\taulest)}{\sqrt{V_{\tau, \ell}^{\text{lim}}}}z_{\alpha/2} - \frac{\sqrt{N}\taul^*}{\sqrt{V_{\tau, \ell}^{\text{lim}}}} \right) - \Phi\left( \frac{\mathrm{S.E.}^{\text{lim}} (\taulest)}{\sqrt{V_{\tau, \ell}^{\text{lim}}}}z_{\alpha/2} + \frac{\sqrt{N}\taul^*}{\sqrt{V_{\tau, \ell}^{\text{lim}}}} \right), \label{eq:power1}
\end{eqnarray}
where $\mathrm{S.E.}^{\text{lim}} (\taulest)$ is the limit of the estimator $\sqrt{N}\widehat{\mathrm{S.E }}(\taulest)$ and $V_{\tau, \ell}^{\text{lim}}$ is the asymptotic limit of the true variance $N\text{Var}(\taulest)$. See Appendix \ref{ss:Appendix3} for a proof of (\ref{eq:power1}).
As noted in \cite{branson2022power}, the limit of the standard error estimator is less than or equal to the (limit of the) true standard error, with the relative magnitude dependent on the amount of treatment effect heterogeneity. 
Thus, power will be reduced if there is more treatment effect heterogeneity, leading to a more conservative variance estimator.

In general settings, treatment effect heterogeneity is not identifiable. Therefore, we propose a strategy that gives conservative approximations based on an upper bound for variance, removing the unidentified part.
Using this strategy, we obtain following approximation for power which gives an upper bound:
\begin{eqnarray}
 \beta \left(\taul^* \right) 
 &\approx&  2 - \Phi\left( z_{\alpha/2} - \frac{\sqrt{N}\taul^*}{\mathrm{S.E}.^{\text{lim}}  (\taulest)} \right) - \Phi\left( z_{\alpha/2} + \frac{\sqrt{N}\taul^*}{\mathrm{S.E}.^{\text{lim}}   (\taulest)} \right).\label{eq:power1a}
\end{eqnarray}
Recalling the expression for $\widehat{\mathrm{S.E.(\taulest)}}$ in (\ref{eq:SE}), in practice, we would replace $\mathrm{S.E}.^{\text{lim}}(\taulest)$ with $\sqrt{N} \ \widetilde{\text{S.E.}}(\taulest)$ where
\begin{equation}
\widetilde{\text{S.E.}}(\taulest) =\sqrt{\frac{1}{2^{2(K-1)}}  \sum_{j=1}^{J} \frac{\tilde{S}^2_j}{N_j} }, \label{eq:SEguess}
\end{equation}
and $\tilde{S}^2_j$ is an estimate or guess for the true variance of potential outcomes under treatment $j$, possibly based on the estimated $s^2_j$ from a pilot study. If a balanced design is desired, then $N_j = N/J$.
The approximate power for testing one-sided alternatives $H_A: \tau >0$ and $H_A: \tau <0$ at level $\alpha$ are, respectively,
\[  \beta \left(\taul^* \right) \approx  1 - \Phi\left( z_{\alpha} - \frac{\sqrt{N}\taul^*}{\mathrm{S.E}.^{\text{lim}}  (\taulest)} \right)
\approx 1 - \Phi\left( z_{\alpha} - \frac{\taul^*}{\widetilde{\text{S.E.}}(\taulest)} \right)
\]
and
\[  \beta \left(\taul^* \right) \approx  \Phi\left( z_{1-\alpha} - \frac{\sqrt{N}\taul^*}{\mathrm{S.E}.^{\text{lim}}   (\taulest)} \right)
\approx \Phi\left( z_{1-\alpha} - \frac{\taul^*}{\widetilde{\text{S.E.}}(\taulest)} \right)
\]

Substituting into (\ref{eq:power1a}) $\alpha = 0.05$, $\taur^* = 0.1875$, and $\widetilde{\mathrm{S.E.}}  (\widehat{\tau}_{\text{R}}) = 0.0917$ (based on estimates in our experiment), we obtain $\beta(0.1875) = 0.534$, which means the probability of detecting a race effect as large as the estimated one in the current experiment with the same standard error is 0.534. Similarly, substituting $\taul^* = 0.1042$, the estimated value of the next two largest effects $\taug$ and $\taugi$, we obtain $\beta(0.104) = 0.206$.

To illustrate computation of the power curve, assuming that the true proportions of the responses $P_1, \ldots, P_8$ in treatment groups are equal to the estimated proportions $p_1, \ldots, p_8$ shown in Table \ref{tab:lawyer_outcomes1}, we compute the power function $\beta \left(\taul^* \right)$ for $\taul^* = 0.1875$ and $\taul^* = 0.1042$ using (\ref{eq:power1a}) corresponding to different values of the sample size $N$ assuming a balanced allocation of $r = N/8$ units into the 8 treatment groups. Substituting the sample variance $s_j^2 = r_0/(r_0-1) \ p_j (1-p_j)$, where $r_0$ is the sample size for each arm in the pilot study, the unbiased estimator of $S_j^2$ in (\ref{eq:SEguess}) based on the pilot data, we adjust our best guess of the (conservative) standard error based on the pilot data as follows:
\[\widetilde{\text{S.E.}}(\taulest) =\sqrt{\frac{1}{2^{4}}  \sum_{j=1}^{J} \frac{s_j^2}{r} }.\]

\begin{figure}[ht]
\centering 
\caption{Power of two-sided test for detecting current race, gender and gender $\times$ income effects for different $N$ for balanced designs.
Top row is with normal approximation, bottom row is based on finite-population simulations. 
Left plots are for IER control  and right plots are for EER control (right).} 
\label{fig:power1}
\includegraphics[width=16cm]{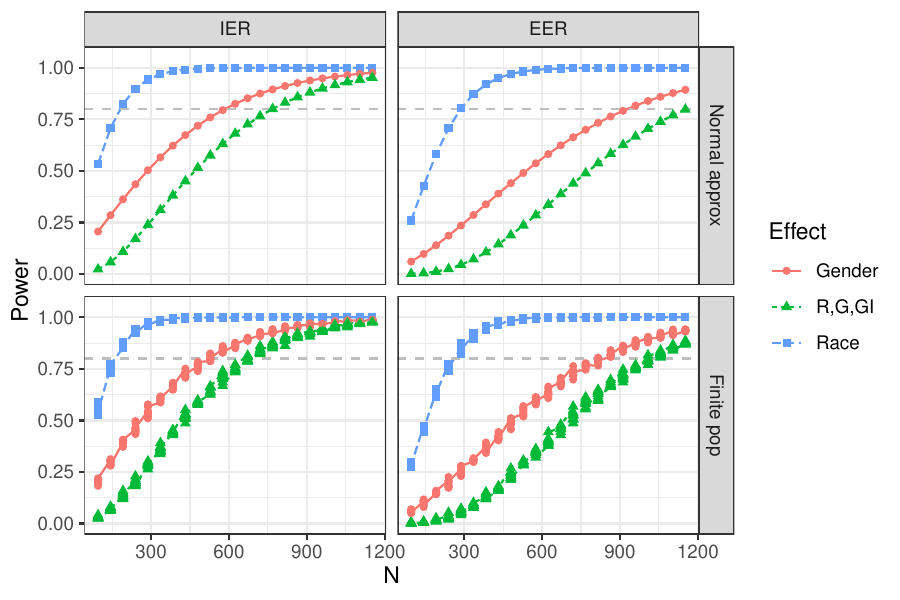}
\end{figure}

The probability of declaring all three effects - main effect of race (assumed to be 0.1875), main effect of gender (assumed to be 0.1042) and interaction effect between gender and income (assumed to be 0.1042) - significant is computed as $\beta(0.1875) \times \left[\beta(0.1042) \right]^2$.
The result is summarized in the left panel of Figure \ref{fig:power1}, which shows that to detect all three factorial effects with a power of 80\% (if they are of the same magnitudes as in the current experiment), one would need a sample size of $N=768$.
This sample size would detect a race effect of magnitude 0.1875 with a probability of almost 1, and a gender effect of 0.1042 with a probability of 0.89.

The power calculations described above are based on the IER-controlled testing procedures, but they can also be based on the Bonferroni EER-controlled procedure described in \ref{ss:IER-EER}. In that case $z_{\alpha/2}$ appearing within both parentheses of (\ref{eq:power1a}) should be replaced by $z_{\alpha/2G}$, where $G$ is the number of independent tests performed. If all factorial effects are assessed simultaneously, then $G = J-1$. The power curves for the EER-controlled procedure obtained exactly in the same manner as those for the IER controlled procedure described above are shown in the right panel of Figure \ref{fig:power1}. 
From the graph, a sample size of $N=1152$ suffices to detect all three factorial effects with approximately 80\% power (if they are of the same magnitudes as in the current experiment) using the EER-controlled procedure.

Figure~\ref{fig:power1} also shows the power analysis under a finite-population.
To generate these, for each sample size a finite-population with potential outcomes matching the stated effect sizes exactly was generated.
Then 10 versions of this finite-population were created by permuting the potential outcomes for each treatment across units, resulting in 10 randomly drawn correlations of potential outcomes.
For each finite-population, random assignment of units to treatments was drawn 1000 times to approximate power.
Each point in the bottom row of Figure~\ref{fig:power1} shows the result for one finite-population.
Based on the average power across the 10 finite-populations for each sample size, a sample size of $N = 720$ and $N=1056$ suffices to get power at least 80\% to detect the two-factor interaction and the two main effects using IER or EER control, respectively.
We see that the results are similar to the asymptotic normal approximation show in the top row of Figure~\ref{fig:power1}, which reassures us that the normal approximation should work well even for finite-population inference.

If we have a desired power level, $\beta$, to detect a particular size of effect for $\taul$, $\taul^*$, then the following Proposition gives an conservative, asymptotic approximation to the necessary sample size for a one-sided test:

\begin{proposition} \label{prop:sample_size}
In a $2^K$ factorial experiment, a conservative, asymptotic approximation to the necessary sample size to ensure that a particular factorial effect $\taul$ with size $\taul^*$ will be declared significant in a size-$\alpha$ one-sided test (with alternative $\taul > 0$) with power $\beta > \max \{\alpha, 0.5 \}$ is
\begin{equation}
N = \left( \text{S.E.}^{\lim}(\taulest) \right)^2 \left(\frac{z_{\alpha}-z_{\beta}}{\taul^*}\right)^2. \label{eq:sample_size}
\end{equation}
\end{proposition}

Proposition \ref{prop:sample_size}, proven in Appendix \ref{ss:Appendix4}, is similar to the result obtained in \cite{branson2022power} and provides a nice way to obtain a desired sample size for factorial experiments with binary outcomes. Recall that for practical purposes, we replaced $\text{S.E.}^{\text{lim}}(\taulest)$ by
\begin{eqnarray*}
\sqrt{N} \ \widetilde{\text{S.E.}}(\taulest) 
= \sqrt{N} \sqrt{\frac{1}{2^{2(K-1)}}  \sum_{j=1}^{J} \frac{\tilde{S}^2_j}{N_j} } 
= \sqrt{\frac{1}{2^{2(K-1)}}  \sum_{j=1}^{J} \frac{\tilde{S}^2_j}{\delta_j} }
\end{eqnarray*}
where $\delta_j = N_j/N$ is the proportion of units desired to be allocated to treatment $j$. Again, recalling that for binary outcomes, $S_j^2 = N/(N-1) P_j (1-P_j)$, we can substitute guesses $\widetilde{P}_j$ for $P_j$ to obtain $\tilde{S}^2_j =  N/(N-1) \widetilde{P}_j (1-\widetilde{P}_j)$, and consequently (\ref{eq:sample_size}) becomes
\[
N = \left(\frac{1}{2^{2(K-1)}}\sum_{j=1}^J \frac{N}{N-1} \frac{\tilde{P}_j(1- \tilde{P}_j)}{\delta_j}\right)\left(\frac{z_{\alpha}-z_{\beta}}{\taul^*}\right)^2
\]

Solving for $N$, we have
\begin{equation}
N = \left(\frac{1}{2^{2(K-1)}}\sum_{j=1}^J\frac{\tilde{P}_j(1- \tilde{P}_j)}{\delta_j}\right)\left(\frac{z_{\alpha}-z_{\beta}}{\taul^*}\right)^2 + 1. \label{eq:sample_size2}
\end{equation}

If we have pilot data, instead of plugging in a guess of $P_j$ directly, we can use the estimates $\tilde{s}_j^2$, the sample variances from the pilot data, which serve as unbiased estimates of $S_j^2$ (at least for the pilot sample).
In this case, we slightly modify the prior equations to obtain
\begin{align}
N = \left(\frac{1}{2^{2(K-1)}}\sum_{j=1}^J  \frac{s_j^2}{\delta_j}\right)\left(\frac{z_{\alpha}-z_{\beta}}{\taul^*}\right)^2.\label{eq:sample_size3}
\end{align}

For example, if we wish to use a balanced design to detect a main effect of size $0.1$, similar to that estimated for gender, at the $\alpha = 0.05$ level with power 0.9 then we would require sample size
\[N \approx \left(\frac{8}{2^4}\sum_{j=1}^8s_j^2\right)\left(\frac{z_{0.05} - z_{0.9}}{0.1}\right)^2 
=690.93.\]
Rounding up, we need at least 691 total experimental units.

\subsection{Power and sample size calculations under optimal allocations} \label{ss:power-optimal}

In the power calculations done so far we focused on a balanced design with $r$ units assigned to each treatment combination so that $N = Jr$.
However, balanced designs are not necessarily optimal with respect to inferential properties. \cite{ravichandran2023optimal} derived results on optimal allocations of $N$ experimental units into $J=2^K$ treatment groups in a completely randomized $2^K$ factorial experiment with respect to different optimality criteria.
These criteria are defined as different functionals of the estimable part of the covariance matrix $\mathbf{V}_{\bm \tau}$ defined in (\ref{eq:cov_matrix}), i.e.,
\begin{equation}
\widetilde{\mathbf{V}}_{\bm \tau} = \frac{1}{2^{2(K-1)}} \sum_{j=1}^{J} \frac{S^2_j}{N_j} \widetilde{\bm  \lambda}_j \widetilde{\bm \lambda_j} ^{\T}. \nonumber 
\end{equation}

Specifically, in \cite{ravichandran2023optimal} three optimality criteria - the D-optimality criterion based on the determinant of $\widetilde{\mathbf{V}}_{\bm \tau}$, the A-optimality criterion based on the trace of $\widetilde{\mathbf{V}}_{\bm \tau}$, and the E-optimality criterion based on the largest eigenvector of $\widetilde{\mathbf{V}}_{\bm \tau}$ - were considered, and the allocation that minimizes each criterion was obtained in terms of $\xi_j = N_j/N$, the proportion of units receiving treatment $j = 1, \ldots, J$.
Based on the results of \cite{ravichandran2023optimal}, under a completely randomized design the A-optimal and E-optimal allocations make $\xi_j$ proportional to $S_j$ and $S^2_j(= \frac{N}{N-1}P_j(1-P_j)$ respectively, whereas the D-optimal allocation is a balanced design.
Because $S_j^2$ are typically unknown, we can use plug-in $\tilde{S}_j^2$ based on guesses of $P_j$ or pilot data.

Thus, the power calculations done earlier will not change under D-optimality. However, for A- and E-optimal allocations, we re-generate power curves similar to the ones in (\ref{fig:power1}) with different sample sizes $N$ by repeating the following steps:
\begin{enumerate}
\item Fix $N$.
\item Determine the optimal $N_1, \ldots, N_J$ (taking $N_j \propto S_j$ for A-optimal allocation and $N_j \propto S^2_j$ for E-optimal allocation).
\item Calculate $\widetilde{\text{S.E.}}(\taulest)$ (for $\ell \in \{1,\dots,J-1\}$) using (\ref{eq:SEguess}).
\item Calculate $\beta\left(\taul^* \right)$ using (\ref{eq:power1a}) for the IER-controlled procedure and replacing $\alpha/2$ by $\alpha/(2J-2)$ in (\ref{eq:power1a}) for the EER-controlled procedure, plugging in our estimates $\widetilde{\text{S.E.}}(\taulest)$ in place of the true variances.
\end{enumerate}

\begin{figure}[ht]
\centering 
\caption{Power of detecting current race, gender and gender $\times$ income effects for different $N$ for A-optimal designs.
Top row is with normal approximation, bottom row is based on finite-population simulations. 
Left plots are for IER control  and right plots are for EER control (right).} \label{fig:power2}
\begin{tabular}{cc}
\includegraphics[width=16cm]{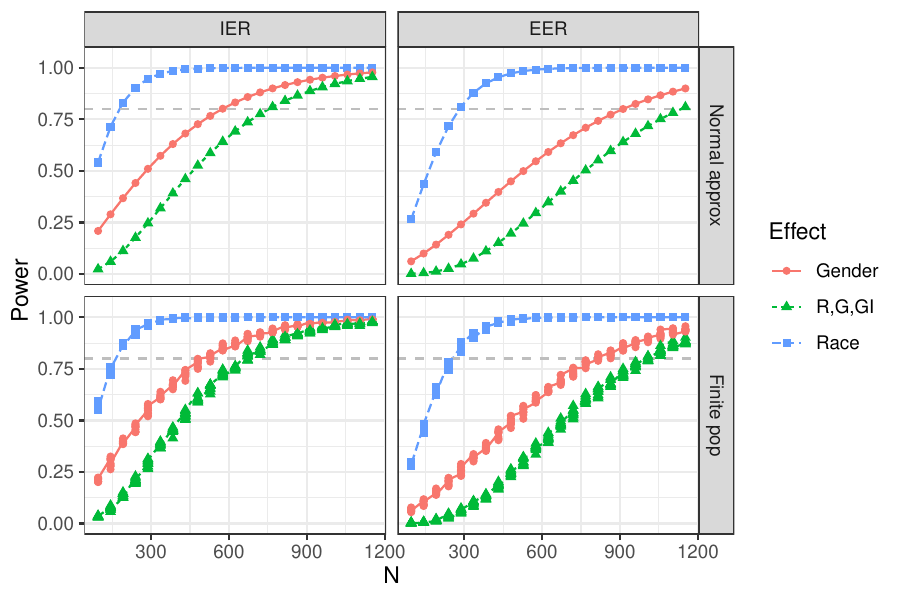}
\end{tabular}
\end{figure}

The results for the A-optimal allocation are summarized in Figure \ref{fig:power2}.
Using the normal approximation, in order to detect all three factorial effects (race, gender and gender $\times$ income) with a power of 80\% if they are of the same magnitudes as in the current experiment, one would need a sample size of $N=768$ under the IER-controlled procedure and $N=1152$ for the EER-controlled procedure respectively.
The results for the E-optimal allocation are similar, and are therefore not shown separately.

The bottom row of Figure~\ref{fig:power2} shows the power analysis under a finite-population, generated in the same way as the previous section.
Based on the average power across the 10 finite-populations for each sample size, we would need a sample size of at least $N = 672$ and $N=1056$ to get power at least 80\% to detect the three-factor interaction using IER or EER control, respectively.
Overall the results are very similar to the asymptotic analysis, though result in a somewhat lower sample size which reflects the impact of treatment effect heterogeneity in the finite-population.
In this case, we see modest sample size savings using optimal allocation.

To conduct sample size calculations without searching over a range of values, we can pick an optimality criteria and use results in \cite{ravichandran2023optimal} to find the optimal proportion of units to assign to each treatment arm, $\delta_j$.
Then we can apply Equation~\ref{eq:sample_size} with the desired power level $\beta$ to find the necessary sample size under our fixed optimal design.

\section{Inference for non-linear causal estimands} \label{sec:other_estimands}

So far, we have discussed inference of finite-population level factorial effects that are linear contrasts of the proportions $P_j$'s (e.g., proportions of lawyers who would potentially respond to an email based on treatment combination $j$). Recall from (\ref{eq:tau_R}) that the main effect of race was defined as
$$ \taur = \frac{1}{4} \lambdar^{\T} \mathbf{P} = \frac{1}{4} \sum_{j=1}^8 \lambdarj P_j,$$
where $\lambdarj$ is the $j$th element of contrast vector $\lambdar = (-1, \ -1, \ -1, \ -1, \ +1, \ +1, \ +1, \ +1)^{\T}$,
so that
$$ \taur = -\frac{P_1 + P_2 + P_3 + P_4}{4}+ \frac{P_5 + P_6 + P_7 + P_8}{4}.$$

Instead of comparing the arithmetic means of the proportions of responses for two race levels, one can also compare the arithmetic means of the logarithms of the proportions, i.e.,
\begin{eqnarray}
\etar &=& \frac{1}{4} \sum_{j=1}^8 \lambdarj \log P_j \label{eq:eta_R} \\
&=& \frac{1}{4} \left( \sum_{j: \lambdarj=+1} \log P_j - \sum_{j: \lambdarj=-1} \log P_j \right) = \log \frac{ \left(\prod_{j: \lambdarj=+1} P_j \right)^{1/4}}{ \left(\prod_{j: \lambdarj=-1} P_j \right)^{1/4}}. \nonumber 
\end{eqnarray}
This is the logarithm of the ratio of the geometric means of proportion of responses in the two racial groups. 
We will call this \emph{new estimand for $2^K$ factorial designs with binary responses} as the \emph{logarithmic factorial effect} (logFE) of race. It is worthwhile to note that for a one-factor experiment at two levels $-1$ and $+1$, the causal estimand reduces to the causal risk ratio (CRR) $\log P_{+1} - \log P_{-1}$ studied in \cite{dingdasgupta2016}. 
Thus logFEs can be considered extensions of the CRR to the case of $2^K$ factorial experiments.

Similar to $\etar$, one can also define a factorial causal estimand replacing the $P_j$'s in $\taur$ or the $\log(P_j)$'s in $\etar$ by the logit transformations of $P_j$'s. i.e., $\log(P_j / (1-P_j))$. For example, the factorial effect of race in the logit scale can be defined as:
\begin{eqnarray}
\thetar = \frac{1}{4} \sum_{j=1}^8 \lambdarj \log \frac{P_j}{1-P_j}. \label{eq:theta_R}
\end{eqnarray}
This causal estimand is an extension of the causal odds ratio (COR) given by
$\log P_{+1} / (1- P_{+1}) - \log P_{-1} / (1- P_{-1})$ studied for a one-factor experiment at two levels $-1$ and $+1$ in \cite{dingdasgupta2016}. We will call this effect, another new estimand for $2^K$ factorial experiments with binary response, the \emph{logit factorial effect} (logitFE) of race.

Plug-in point estimates of $\etar$ and $\thetar$ can be obtained by substituting the observed proportions $p_j$ in place of $P_j$ in (\ref{eq:eta_R}) and (\ref{eq:theta_R}). These estimators are biased, and as discussed in \cite{dingdasgupta2016}, the absence of linearity makes exact variance calculations intractable. However, approximate inference can be made using their asymptotic distributions that can be derived using Taylor series expansions and the finite-population delta method \citep{pashley2022note}. We state two results, proven in Appendix \ref{ss:Appendix5}, that extend the asymptotic distributions of the plug-in estimators of CRR and COR in \cite{dingdasgupta2016} to the case of completely randomized $2^K$ factorial experiments with binary responses.

\begin{proposition} \label{Prop2}
For $J = 2^K$ and $j=1, \ldots, J-1$, let $$\eta_{\ell} = \frac{1}{2^{K-1}} \sum_{j=1}^J \lambda_{\ell j} \log P_j$$ denote the $\ell$th logFE and
$$\widehat{\eta}_{\ell} = \frac{1}{2^{K-1}} \sum_{j=1}^J \lambda_{\ell j} \log p_j$$ denote its plug-in estimator, where $P_j$ and $p_j$ are respectively the true and observed proportion of responses for treatment $j$ and $(\lambda_{\ell 1}, \ldots, \lambda_{\ell J})^{\T}$ is the contrast vector associated with the factorial effect. Then, under the conditions discussed in the Appendix, $\widehat{\eta}_{\ell}$ is consistent for $\eta_{\ell}$, and is asymptotically normal with variance
\begin{equation}
\frac{1}{2^{2(K-1)}} \left( \sum_{j=1}^J \frac{(N-N_j)S_j^2}{N N_j P_j^2 } - \frac{1}{N} \sum_{j=1}^{J-1} \sum_{j^{\prime}={j+1}}^J  \lambda_{\ell j} \lambda_{\ell j^{\prime}} \frac{S_j^2 + S_{j^{\prime}}^2 - S^2_{j - j^{\prime}}}{P_j P_{j^{\prime}} } \right), \label{eq:varetahat}
\end{equation}
where $S_j^2$ is the variance of potential outcomes for treatment $j$ defined in (\ref{eq:Ssqj}), and 
$$S^2_{j - j^{\prime}} = \frac{1}{N-1} \sum_{i=1}^N \left( Y_i(j) - Y_i(j^{\prime}) -P_j - P_{j^{\prime}} \right)^2 $$
is the variance of unit-level differences of potential outcomes between treatments $j$ and $j^{\prime}$.
\end{proposition}

\begin{proposition} \label{Prop3}
For $J = 2^K$ and $j=1, \ldots, J-1$, let $$\theta_{\ell} = \frac{1}{2^{K-1}} \sum_{j=1}^J \lambda_{\ell j} \log \frac{P_j}{1-P_j}$$ denote the $\ell$th logitFE and
$$\widehat{\theta}_{\ell} = \frac{1}{2^{K-1}} \sum_{j=1}^J \lambda_{\ell j} \log \frac{p_j}{1-p_j}$$ denote its plug-in estimator, where $P_j$ and $p_j$ are respectively the true and observed proportions of responses for treatment $j$ and $(\lambda_{\ell 1}, \ldots, \lambda_{\ell J})^{\T}$ is the contrast vector associated with the factorial effect. Then, under analogous conditions to Proposition~\ref{Prop2}, $\widehat{\theta}_{\ell}$ is consistent for $\theta_{\ell}$, and is asymptotically normal with variance
\begin{equation}
\frac{1}{2^{2(K-1)}} \left( \sum_{j=1}^J \frac{(N-N_j)S_j^2}{N N_j P_j^2 (1-P_j)^2} - \frac{1}{N} \sum_{j=1}^{J-1} \sum_{j^{\prime}=j+1}^J \lambda_{\ell j} \lambda_{\ell j^{\prime}} \frac{S_j^2 + S_{j^{\prime}}^2 - S^2_{j - j^{\prime}}}{P_j P_{j^{\prime}} (1 - P_j) (1 - P_{j^{\prime}}) } \right), \label{eq:varthetahat}
\end{equation}
where $S_j^2$ is defined in (\ref{eq:Ssqj}), and 
$S^2_{j - j^{\prime}}$ is as defined in Proposition~\ref{Prop2}.
\end{proposition}

The sampling variances of $\widehat{\eta}_{\ell}$ and $\widehat{\theta}_{\ell}$ can be respectively estimated by substituting the sample variances $s_j^2$ for $S_j^2$, $p_j$ for $P_j$ and ignoring the non-estimable term $S^2_{j - j^{\prime}}$ in (\ref{eq:varetahat}) and (\ref{eq:varthetahat}). Similar to the inference for $\taulest$, these variance estimators will asymptotically have positive bias in finite-population settings unless strict additivity holds, making the resulting inference asymptotically conservative. In the current study, the point estimates of the logFE and logitFE of race are obtained as $\widehat{\eta}_{\textrm{R}} = 0.63$ and $\widehat{\theta}_{\textrm{R}} = 0.91$. One-sided and two-sided 95\% confidence intervals for $\eta_{\textrm{R}}$ are [0, 1.85] and [-0.82, 2.08] respectively, and one-sided and two-sided 95\% confidence intervals for $\theta_{\textrm{R}}$ are [0, 2.13] and [-0.54, 2.36] respectively.


\section{Discussion and concluding remarks} \label{sec:discussion}

In this paper, we have laid out an extensive methodology for conducting randomization-based Neymanian inference of causal effects of signals of attributes on a binary response variable in audit experiments with a $2^K$ factorial structure. The methodology has been motivated, expounded, and demonstrated using a recently reported discrimination experiment. We have used and extended available methodology and tools used in recent literature on Neymanian inference of experiments with binary responses, our specific contributions being developing methods for power calculations and sample size determination under balanced and possibly unbalanced, optimal design settings, extending definitions of non-linear causal estimands from two-armed experiments to the factorial setting, and proposing asymptotic Neymanian inference of such estimands.

There are a few directions related to causal inference from randomized $2^K$ experiments with binary outcomes that can be explored further.
As noted at the end of Section \ref{ss:POT}, an inferential procedure similar to the one adopted for the finite population of 96 lawyers in the \cite{Libgober2020} study can also be applied for inference of estimands defined with respect to a super population (e.g., all lawyers in the state of California) from which the 96 lawyers can be assumed to be randomly sampled. 
The point and interval estimators for the finite-population estimands and super-population estimands would remain the same; however inference for the super population will not be as conservative as the finite-population inference.
A better way to carry out separate analyses for the finite and super populations is to use a Bayesian approach. Bayesian model-based analysis of factorial designs is a natural extension of randomization-based analysis to incorporate uncertainty into the potential outcomes. \cite{DPR2015} demonstrated application of this framework using a normal hierarchical model with a compound symmetric covariance matrix, normal prior on the mean vector and an inverse gamma prior on the variance. For binary outcomes, \cite{dingdasgupta2016} developed a procedure for Bayesian inference for linear and non-linear causal effects assuming independent potential outcomes under treatment and control and also proposed a sensitivity analysis with respect to the unknown association parameter. Extending such an analysis to the case of $2^K$ factorial designs will be a useful addition to the existing literature.


\bigskip

\noindent \textbf{Acknowledgement}: This research was partially supported by National Science Foundation grant SES 2217522.
Any opinions, findings, conclusions, or recommendations expressed in this material are those of the authors and do not necessarily reflect the views of the National Science Foundation.


\appendix
\section{Appendix} \label{sec:Appendix}

\subsection{Conditions for asymptotic Normality} \label{ss:Appendix1}

From Theorem 5 of \cite{LiDingCLT2017}, if the following conditions hold:
\begin{enumerate}
\item $N_j/N$ has a positive limiting value
\item $S^2_j$ and $S_{jj'}$ have positive limiting values, where $S_{jj'} = \frac{1}{N-1}\sum_{i=1}^N(Y_i(j) - P_j)(Y_i(j') - P_{j'})$
\item $\max_{j \in \{1,\dots,J\}}\max_{i \in \{1,\dots,N\}}(Y_i(j) - P_j)^2/N \to 0$
\end{enumerate}
then $N\mathbf{V}_{\bm \tau}$ has a limiting value, which we will denote $\mathbf{V}_{\bm \tau}^{\text{lim}}$ and 
\[\sqrt{N}(\bm {\widehat{\tau}} - \bm {\tau}) \overset{d}{\to} \text{N}(\bm{0}, \mathbf{V}_{\bm \tau}^{\text{lim}}). \]

Note that the third condition is automatically satisfied and the second will also be automatically satisfied as long as not all units have the same potential outcomes for each treatment condition (e.g., no $j$ such that $S^2_j = 0$), due to the binary nature of the potential outcomes.

\subsection{Conservativeness of p-values} \label{ss:Appendix2}

Let $\mathrm{S.E.}^{\text{lim}}(\taulest)$ be the limiting value of $\sqrt{N}\widehat{\mathrm{S.E.}}(\taulest)$.
Let $V_{\tau, \ell}^{\text{lim}}$ be the $\ell +1$ diagonal entry of $\mathbf{V}_{\bm \tau}^{\text{lim}}$.

We have
\begin{eqnarray*}
P\left(\frac{\taulest}{\widehat{\mathrm{S.E.}}(\taulest)} \geq c \right) &=& P\left(\frac{\sqrt{N}\taulest}{\sqrt{V_{\tau, \ell}^{\text{lim}}}}  \geq c\frac{\sqrt{N}\widehat{\mathrm{S.E.}}(\taulest)}{\sqrt{V_{\tau, \ell}^{\text{lim}}}} \right) \\ 
&\approx& P\left(\frac{\sqrt{N}\taulest}{\sqrt{V_{\tau, \ell}^{\text{lim}}}}  \geq c\frac{\mathrm{S.E.}^{\text{lim}}(\taulest)}{\sqrt{V_{\tau, \ell}^{\text{lim}}}} \right), \ \mbox{by Appendix \ref{ss:Appendix1}} \\
&=& 1 - \phi\left(c\frac{\mathrm{S.E.}^{\text{lim}}(\taulest)}{\sqrt{V_{\tau, k}^{\text{lim}}}} \right) \leq 1 - \phi\left(c \right),
\end{eqnarray*}
for $c>0$.
The last line is because $\mathrm{S.E.}^{\text{lim}}(\taulest) \geq \sqrt{V_{\tau, \ell}^{\text{lim}}}$.
The magnitude of $\mathrm{S.E.}^{\text{lim}}(\taulest)/ V_{\tau, \ell}^{\text{lim}}$ depends on the (relative) size of the (limit of the) treatment effect heterogeneity term, $S^2_{\tau,\ell}$.

\subsection{Proof of (\ref{eq:power1})} \label{ss:Appendix3}

\begin{eqnarray*}
 \beta \left(\taul^* \right) &=& \Pr \left( |T_{\ell}| \ge z_{\alpha/2} | \ \taul = \taul^* 
 \right) = 1 - \Pr \left( \left| \frac{\taulest}{\widehat{\mathrm{S.E.}} (\taulest)} \right| \le z_{\alpha/2} \ \Big| \ \taul = \taul^* \right)  \\
  &\approx& 1 - \Pr \left( \left| \frac{\sqrt{N}\taulest}{\mathrm{S.E.}^{\text{lim}} (\taulest)} \right| \le z_{\alpha/2} \ \Big| \ \taul = \taul^* \right)  \nonumber \\
  &=& 1 - \Pr \left(z_{1-\alpha/2}  \leq\frac{\sqrt{N}\taulest}{\mathrm{S.E.}^{\text{lim}} (\taulest)}  \le z_{\alpha/2} \ \Big| \ \taul = \taul^* \right)  \\
  &=& 2 - \Phi\left( \frac{\mathrm{S.E.}^{\text{lim}} (\taulest)}{\sqrt{V_{\tau, \ell}^{\text{lim}}}}z_{\alpha/2} - \frac{\sqrt{N}\taul^*}{\sqrt{V_{\tau, \ell}^{\text{lim}}}} \right) - \Phi\left( \frac{\mathrm{S.E.}^{\text{lim}} (\taulest)}{\sqrt{V_{\tau, \ell}^{\text{lim}}}}z_{\alpha/2} + \frac{\sqrt{N}\taul^*}{\sqrt{V_{\tau, \ell}^{\text{lim}}}} \right).
\end{eqnarray*}

\subsection{Proof of Proposition \ref{prop:sample_size}} \label{ss:Appendix4}

Consider testing $H_0: \taul =0$ vs $H_A: \taul > 0$.
We want power $\beta$ against the alternative value $\taul = \taul^* >0$.

That is, we want
\begin{align*}
\beta &=\mathrm{pr}\left(\frac{\sqrt{N}(\taulest - \taul^*)}{\text{S.E.}^{\text{lim}}(\taulest)} > z_{\alpha} - \frac{\sqrt{N}\taul^*}{\text{S.E.}^{\text{lim}}(\taulest)} \Big| \taul = \taul^*\right)\\
& = 
\mathrm{pr}\left(\frac{\sqrt{N}(\taulest - \taul^*)}{V_{\tau,\ell}^{\text{lim}}} > \frac{\text{S.E.}^{\text{lim}}(\taulest)}{V_{\tau,\ell}^{\text{lim}}}\left[z_{\alpha} - \frac{\sqrt{N}\taul^*}{\text{S.E.}^{\text{lim}}(\taulest)}\right] \Big| \taul = \taul^*\right)\\
&=1-\Phi\left(\frac{\text{S.E.}^{\text{lim}}(\taulest)}{V_{\tau,\ell}^{\text{lim}}}\left[z_{\alpha} - \frac{\sqrt{N}\taul^*}{\text{S.E.}^{\text{lim}}(\taulest)}\right] \right).
\end{align*}

Simplifying, we obtain
\begin{align*}&\frac{\text{S.E.}^{\text{lim}}(\taulest)}{V_{\tau,\ell}^{\text{lim}}}\left[z_{\alpha} - \frac{\sqrt{N}\taul^*}{\text{S.E.}^{\text{lim}}(\taulest)}\right] = z_{\beta}, \\
&\Rightarrow   \frac{\sqrt{N}\taul^*}{\text{S.E.}^{\text{lim}}(\taulest)} = z_{\alpha}-\frac{V_{\tau,\ell}^{\text{lim}}}{\text{S.E.}^{\text{lim}}(\taulest)}z_{\beta}\\
&\Rightarrow N = \left(\frac{\text{S.E.}^{\text{lim}}(\taulest)z_{\alpha}-V_{\tau,\ell}^{\text{lim}}z_{\beta}}{\taul^*}\right)^2
\end{align*}
This is very similar to the expression derived in \cite{branson2022power} for two-armed experiments.

If $\beta > \alpha$ and $\beta > 0.5$, then $z_{\alpha} > -z_{\beta} > 0$ and since $\text{S.E.}^{\text{lim}}(\taulest)>V_{\tau,\ell}^{\text{lim}}$, we must have
\begin{align*}
\left(\frac{\text{S.E.}^{\text{lim}}(\taulest)z_{\alpha}-V_{\tau,\ell}^{\text{lim}}z_{\beta}}{\taul^*}\right)^2 \leq 
(\text{S.E.}^{\text{lim}}(\taulest))^2\left(\frac{z_{\alpha}-z_{\beta}}{\taul^*}\right)^2.
\end{align*}
Thus, setting $N = (\text{S.E.}^{\text{lim}}(\taulest))^2\left(\frac{z_{\alpha}-z_{\beta}}{\taul^*}\right)^2$ is conservative. In practice, we would plug in an estimate of $\text{S.E.}^{\text{lim}}(\taulest)$ of the form
\[ \sqrt{\frac{1}{2^{2(K-1)}}  \sum_{j=1}^{J} \frac{\tilde{S}^2_j}{\delta_j} }.\]

\subsection{Proof of Propositions \ref{Prop2}} \label{ss:Appendix5}

Here we consider a completely randomized $2^K$ factorial experiment, with $J=2^K$ treatment combinations indexed by $j = 1, \ldots, J$.  Let $P_j$ and $p_j$ denote, respectively, the true and observed proportions of responses for treatment $j$. We first state the following Lemma on the sampling distribution of $p_j$, the proof of which follows directly from \cite{dingdasgupta2016} (Supplementary materials).
\begin{lemma} \label{lemma1}
The sampling distribution of $p_j$'s satisfy the following:
\begin{eqnarray*}
\E(p_j) &=& P_j, \quad \var(p_j) = \frac{N-N_j}{N}.\frac{S_j^2}{N} \\
\cov(p_j, p_{j^{\prime}}) &=& - \frac{1}{2N} \left( S_j^2 + S^2_{j^{\prime}} - S^2_{j - j^{\prime}} \right), \ j \ne j^{\prime} 
\end{eqnarray*}
\end{lemma}

Under the previously stated conditions for asymptotic normality and an assumption that $P_j$ is bounded away from 0 for all $j$ to ensure continuity of the estimand, \cite{pashley2022note} gives us that $\widehat{\eta}_{\ell}$ is asymptotically normal with variance
\begin{align*}
    \var(\widehat{\eta}_{\ell}) &= 
    \frac{1}{2^{2(K-1)}}\sum_{j=1}^J\frac{S^2_j}{p_j^2(1-p_j)^2N_j} - \frac{1}{N(N-1)} \frac{1}{2^{2(K-1)}}\sum_{i=1}^N\left(\sum_{j=1}^J\lambda_{\ell j} \frac{Y_i(j) - P_j}{P_j(1-P_j)}\right)^2.
\end{align*}
The last term is not identifiable and can be dropped to make an asymptotically conservative estimator.
We can also write the asymptotic variance as
\begin{eqnarray*}
\var(\widehat{\eta}_{\ell}) 
&=& \frac{1}{2^{2(K-1)}} \left( \sum_{j=1}^J \frac{(N-N_j)S_j^2}{N N_j P_j^2 } - \frac{1}{N} \sum_{j=1}^{J-1} \sum_{j^{\prime}={j+1}}^J  \lambda_{\ell j} \lambda_{\ell j^{\prime}} \frac{S_j^2 + S_{j^{\prime}}^2 - S^2_{j - j^{\prime}}}{P_j P_{j^{\prime}} } \right)
\end{eqnarray*}

\medskip

The Proof of Proposition \ref{Prop3} is similar and is skipped.


\bibliographystyle{apalike}
\bibliography{samplesize}

\end{document}